\documentclass[twocolumn,amsmath,showpacs,nofootinbib]{revtex4-2}
\usepackage{multirow}
\usepackage{amssymb}
\usepackage{amsmath, graphicx}
\usepackage{dcolumn}
\usepackage{bm}
\usepackage{enumerate}
\usepackage{slashed}
\usepackage{epstopdf}
\usepackage[unicode]{hyperref}
\usepackage{csquotes}
\usepackage[usenames]{color}

\newcommand{\be}{\begin{equation}}
\newcommand{\ee}{\end{equation}}
\newcommand{\bea}{\begin{eqnarray}}
\newcommand{\eea}{\end{eqnarray}}
\newcommand{\ben}{\begin{enumerate}}
\newcommand{\een}{\end{enumerate}}
\newcommand{\bde}{\begin{widetext}}
\newcommand{\ede}{\end{widetext}}

\newcommand{\al}{\alpha}
\newcommand{\la}{\lambda}

\newcommand{\bc}{\begin{center}}
\newcommand{\ec}{\end{center}}

\newcommand{\AdrHEPC}{$^a$Department of Theoretical Physics, Faculty of Physics and Engineering Physics,\\ University of Science, Ho Chi Minh City, Vietnam\\ $^b$Vietnam National University, Ho Chi Minh City, Vietnam}
\usepackage{braket}

\begin{document}

\title{\boldmath Inflation with the standard and Randall–Sundrum model in the Two-time Physics}

\author{Vo Quoc Phong$^{a,b}$}
\email{vqphong@hcmus.edu.vn}
\affiliation{\AdrHEPC}
\date{\today}	

\begin{abstract}
We propose a scalar inflationary potential as  $V(\phi)=M^4\phi^{2n-2}(\phi^{2n}+m^{2n})^{1/n-1}$. This potential is similar to the shaft inflation one. However,  they satisfy the $Z_2$ symmetry for all $n$. The potential may come from the Higgs-dilaton potential in the two-time (2T) physics. The slow-roll scenario is recomputed in the 4-dimension (4D) and Randall-Sundrum II (RSII) frameworks. The tensor-to-scalar ratio in the RSII model is always higher than in the 4D model and is in good agreement with the experimental data of BICEP2 and Planck. Comparing this with Planck data, we estimate $M_5$ to be around $[1-2]\times 10^{16}$ GeV. Furthermore, the potential allows much lower scalar field exponents than other potentials, which results in high agreement with experimental data. Moreover, the results also reinforce the models that have the extra dimensions, should be focused. The inflation data can be used to test for the existence of the extra dimensions.
\end{abstract}
\maketitle

\tableofcontents

\section{Introduction}

The inflation is the dominant paradigm to handle the fine-tuned challenge of the Universe's early conditions \cite{guth}. As for the nature of inflaton, there are many hypotheses. It can be summarized as follows: Inflaton can be Higgs, an exotic scalar particle, or one particle outside the SM. The search for inflationary models is a vibrant area of research, providing a wide range of potentials to trace the causes as well as assumptions about exotic particles or new physics. There are two basic types of inflation today: the first is the slow-roll form and the second is the non-slow-roll forms like the warm inflation \cite{warm,warm2}. However, the concave-like inflaton potentials are more advantageous, as shown by the most recent findings from Planck, WMAP, and BICEP/Keck observations, which indicate that there are no large primordial tensor perturbations \cite{bicep}. 

The hierarchy problem—the enormous difference between the electroweak and Planck scales—inspired several solutions by altering the very structure of spacetime. In particular, some models explain the hierarchy problem by adding one or more additional dimensions. The fundamental hint is that gravity is significantly weaker than other forces because gravity may escape through additional dimensions. These extra dimensions are theoretically compatible with quantum gravity in string theory \cite{joe}. They are often expressed as being very small. However, in some models, there are suggestions that their size may still be sufficient for current experimental observations; one of the typical models is the RSII model \cite{EPJC}.

The study of the current inflation problem can be divided into two main approaches: the first one focuses on finding a good inflationary potential, and the second one examines the inflation within frameworks (like $f(R)$, Palatini \cite{ref19} or Einstein). The inflation models can be summarized as the chaotic \cite{chaotic, chaoticb}, warm \cite{warm,warma,warmb}, natural \cite{nifla},  hybrid \cite{hyi, hyib} and the shaft inflation \cite{kostas}, among others, many of which are adjusted in various ways.

However, the nature or origin of the inflation potentials has not yet been discussed in detail. We start from the theoretical framework of the two-time physics (2T physics) \cite{14}, with the dilaton potential. From there, we apply this to the study of inflation, as a step in explaining the feasibility of the inflation potentials.

Current inflation data, although scarce, is sufficient to determine several important quantities such as The tensor-to-scalar ratio $(r)$ and the scalar spectral index $(n_s)$ or the number of e-folds ($N$). These three quantities are now the three main test parameters for any scalar field inflation model. Without exception, in this article, we also focus on determining these parameters.

Therefore, the RSII model \cite{lisa1,lisa2} is currently popular due to its simplicity (having only one extra dimension, which is sufficient to confirm it as an additional dimension) and its ability to meet many expectations, as well as this model's timeliness \cite{maldacena}. We will analyze an inflation scenario of the inflaton potential in the RSII model.

In this paper, we follow the first approach: searching for a scalar-field inflation model and attempting to explain its origins by using the 2T physics. Furthermore, as discussed regarding the RSII model, we will examine new inflation potentials within the RSII framework (they can be reduced from the spacetime of the 2T physics) and compare them with empirical data.

The specific structure of the article is as follows. We propose that a scalar field inflation potential can come from the dilaton potential in the Two-time physics. In this article, by modifying the shaft inflation potential \cite{kostas,kostas2, locjcap}, we propose a new form of potential. We are calculating the slow-roll scenario in the 4 dimensions and the RS II model. In Sec. \ref{2T}, we summarize the reduction of the metric and the dilaton field from the 2T to 1T. And assume that the dilaton is an inflaton. In Sec. \ref{4D} we calculate the new inflation model in the 4D spacetime. In Sec. \ref{RS2} we calculate the inflation in the RSII model. In Sec. \ref{comparisons}, we compare the outcomes of two models with observations and with each other. Sec. \ref{conclu} contains the conclusions. Throughout this article, $\hbar=c=k_B=1$ is utilized to represent natural units.

\section{The dilaton potential}\label{2T}

\subsection{The 2T metric reduction}
The $AdS_6$ space-time of the 2T physics is described by the metric,
\begin{equation}\label{Metric of Hperboloid}
	ds^2=-(dX^0)^2-(dX^5)^2+\sum_{i=1}^4(dX^i)^2.
\end{equation}

The above (4+2) dimensional space-time will be transformed into the (3+1) dimensional space-time with the below transformation (a gauge fixing), 
\begin{align}
	\begin{split}
		X^5&=\frac{1}{2z}(1+z^2(L^2+\overline{x}^2-t^2)),\\
		X^4&=\frac{1}{2z}(1-z^2(L^2-\overline{x}^2+t^2)),\\
		X^\mu&=Lzx^\mu.
	\end{split}
\end{align}

In which $\mu=0,1,2,3$ and $\overline{x}^2\equiv\sum_{i=1}^3(x^i)^2$. The spacetime interval is rewritten as follows

\begin{align}
	ds^2&=\frac{L^2}{z^2}dx^\mu dx_\mu+\frac{L^2}{z^2}dz^2\\
	&=\frac{L^2}{z^2}(\eta_{\mu\nu}dx^\mu dx^\nu+dz^2).
\end{align}
We make another change of variable as
\begin{equation}
	z\rightarrow\frac{1}{r};\hspace{1cm}x^\mu\rightarrow\frac{x^\mu}{L^2}.
\end{equation}
The metric becomes
\begin{equation}
	ds^2=\frac{r^2}{L^2}\eta_{\mu\nu}dx^\mu dx^\nu+\frac{L^2}{r^2}dr^2.
\end{equation}
We make a final change of variable,
\begin{equation}
	r=Le^{\frac{-\sigma}{L}}\rightarrow dr=-e^{\frac{-\sigma}{L}}d\sigma\rightarrow dr^2=e^{\frac{-2\sigma}{L}}d\sigma^2.
\end{equation}
Then the metric becomes
\begin{equation}
	ds^2=e^{\frac{-2\sigma}{L}}\eta_{\mu\nu}dx^\mu dx^\nu+d\sigma^2.
\end{equation}
This is exactly the form of the Randall-Sundrum metric. Accordingly, we have the first-order derivative operator, which can be simplified as follows

\begin{align}
	\begin{split}
		X^M\partial_M&=X^\mu\partial_\mu+X^4\partial_4+X^5\partial_5\\
		&=\frac{x^\mu\partial}{\partial x^\mu}+\frac{(z^2-L^2+\overline{x}^2-t^2)\left(\frac{x^\mu}{L^2}\frac{\partial}{\partial x^\mu}+\frac{z}{L^2}\frac{\partial}{\partial z}\right)}{2}\\
		&-\frac{1}{2}(z^2+L^2+\overline{x}^2-t^2)\left(\frac{x^\mu}{L^2}\frac{\partial}{\partial x^\mu}+\frac{z}{L^2}\frac{\partial}{\partial z}\right)\\
		&=x^\mu\frac{\partial}{\partial x^\mu}-L^2\left(\frac{x^\mu}{L^2}\frac{\partial}{\partial x^\mu}+\frac{z}{L^2}\frac{\partial}{\partial z}\right)\\
		&=-z\frac{\partial}{\partial z}.
	\end{split}
\end{align}
Using the kinematical Klein-Gordon equation for scalar field in the 2T-physics \cite{14,bars1,bars2} in the 4+2 dimensions,
\begin{align}
	\begin{split}
		(X^M\partial_M+1)\Phi&=0.\\
		\left(z\frac{\partial}{\partial z}-1\right)\Phi&=0.
	\end{split}
\end{align}
One solution to the above equation is
\begin{equation}
\Phi(X)=z\phi(x).\label{10}
\end{equation}

In Ref. \cite{14}, by another gauge fixing, the authors fitted the dilaton field in 4+2 dimensions to the inflaton field in the 3+1 dimensions as
\begin{equation}
	\Phi(X)=\frac{1}{\kappa}\phi(x).\label{11}
\end{equation}
The dilaton field reduction in Eq.\ref{10} is exactly equivalent to Eq.\ref{11}.

As shown in the above reduction, we can reduce the 2T metric to the standard 4D metric or the metric of the RSII model. And the Higgs-dilaton potentials can also be reduced to the 4D. This implies that we can use the dilaton potential to examine inflation in the 4D and RSII.

\subsection{The Higgs-dilaton potential}

The Higgs-dilaton potential has the following fourth-order form:
\begin{equation}
V(H,\Phi)=f(\mathcal{S}/\Phi)\Phi^4,\label{13}
\end{equation}

where $\mathcal{S}=H^\dagger H$. $H$ is the Higgs doublet. Currently, the Higgs-dilaton potential has been investigated in cases such as those in Table \ref{phanloai}.

\begin{table}[!ht]
	\centering	
	\begin{tabular}{|c|c|}\hline
		$f(\mathcal{S}/\Phi)$&$V(H,\Phi)=f(\mathcal{S}/\Phi)\Phi^4$\\\hline	
		$\frac{\lambda}{4}(\frac{\mathcal{S}^2}{\Phi^2}-\alpha^2)^2+\rho/4$&$\la\left(H^\dagger H - \al ^2 \Phi^2\right)^2+\rho \Phi^4$\\
		\hline		
		$\frac{\lambda}{4}(\frac{\mathcal{S}^2}{\Phi^2}-\alpha^2)^2+\frac{\rho}{4}+\frac{\omega^2}{\kappa^2\Phi^2}$&$\la\left(H^\dagger H - \al ^2 \Phi^2\right)^2+\rho \Phi^4+\frac{\omega^2\Phi^2}{\kappa^2}$\\
		\hline
	\end{tabular}
	\caption{Cases of the Higgs-dilaton potential \cite{phong,phong2025}.}\label{phanloai}
\end{table}

The potential forms in Table \ref{phanloai} have been studied in the warm inflation scenarios \cite{phong} or used for calculating the electroweak phase transition  and gravitational waves \cite{phong2025}.

The potential Eq.\ref{13} has two components, the first is the Higgs component and the interaction between the Higgs and the dilaton, the second is the dilaton component.

In the simplest case, corresponding to the first row in Table \ref{phanloai}, the dilaton potential has the form $V(\Phi)\sim \Phi^4$. After reducing to 1T, $V(\phi)\sim \phi^4$, which corresponds to the warm inflation. However in this paper we will examine more general cases.

If we split $f(\mathcal{S}/\Phi)$ into two components, such that $f(\mathcal{S}/\Phi)=F(\mathcal{S}/\Phi)+G(\Phi)$. The dilaton potential in 2T suggests an inflationary scenario corresponding to a function $G(\Phi)$.

\section{The inflation in the 4D spacetime}\label{4D}

Reducing from the superpotential to the monomial F-term potential \cite{kostas}; the standard shaft inflation model in the 3+1 dimensional spacetime \cite{kostas} is often expressed as
\begin{equation}\label{potentialt}
	\mathcal{V}(\phi)=M^4\phi^{2(n-1)}(\phi^n+m^n)^{2(1-n)/n}.
\end{equation}

The above potentials are not $z_2$ symmetric for odd integers $n$ like $n=3$. So we transform them as follows:
\begin{equation}\label{potential}
V(\phi)=M^4\phi^{2n-2}(\phi^{2n}+m^{2n})^{1/n-1}.
\end{equation}

When $\phi \ll m$ the potential becomes monomial, but when $\phi\gg m$ the potential with a constant form or superpotential will be linear in $\phi$. These are similar to those in the shaft inflation.

\begin{figure}[h!]
\centering
\includegraphics[scale=0.8]{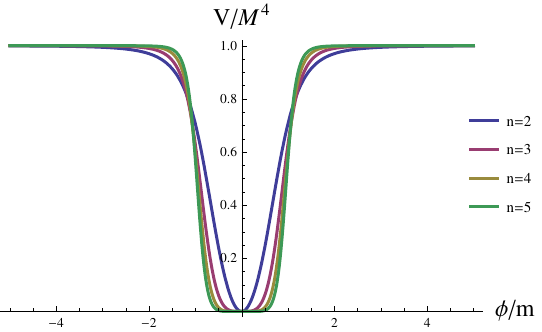}
\caption{The inflaton potential with $n=2,3,4,5$.}
\label{potential_figure}
\end{figure}

Comparing the potential (\ref{potential}) with the dilaton potential in the 2T Physics, it leads to
\begin{equation}\label{potentialf}
	G(\phi)=\phi^{2n-6}(\phi^{2n}+m^{2n})^{1/n-1},
\end{equation}
where $n\geq 2$ is an integer, $M$ is the energy scale of inflation, and $m$ is about the threshold energy scale for the inflaton field to halt slow-roll (SR). The key characteristic is that the potential resembles a monomial chaotic inflation when $\phi\ll m$, but it approaches a constant value (plateau) when $\phi\gg m$. Figure \ref{potential_figure} shows the plot of the potential. Since the inflation often occurs when the inflaton field is above this energy scale in large-field inflation models, we will use $m\simeq 10^{18}$ GeV (for a study of shaft inflaton potential in the warm inflation scenario, see \cite{gas, locjcap} as well). Additionally, the warm inflation scenario in the Two-time physics has also been studied in Ref.\cite{phong}.

The potential in Ref.\cite{kostas} does not have the $Z_2$ symmetry with $\phi$ when $n$ is odd but the potential Eq.\ref{potential}, there is always this symmetry for every $n$. For any real $n$ as the lines in Fig.\ref{potential_figure}, the above potential always gives a SR one.

In Eq. \ref{potentialf}, when $\phi \ll m$ and $n=3$, $G(\phi)\sim const$, so $V(\phi)\sim \phi^4$. This is very much in line with the dilaton potential in the 2T Physics.

Furthermore, when $\phi \gg m$ and $n=3$, $G(\phi)\sim \phi^{-4}$ and $V(\phi)\sim const$. This is also consistent with the dilaton potential in the 2T Physics. In the last section, we will evaluate the case $n=3$ in detail. 

Next we will use Eq. \ref{potential} to examine inflation in the 4D case in this section and the RSII model in Sec.\ref{RS2}.

\begin{figure}[h!]
\centering
\includegraphics[scale=0.8]{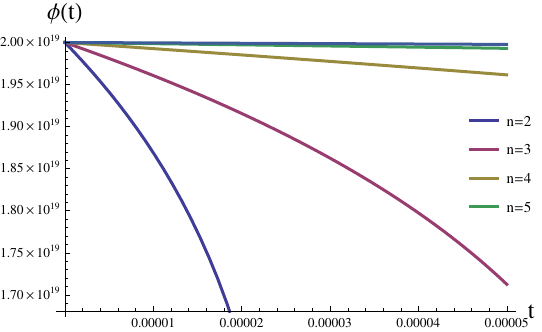}
\caption{The $\phi(t)$ field in the standard 4D spacetime model with $n=2,3,4,5$, $M\simeq 10^{15}$ GeV, $M_4\simeq 10^{19}$ GeV, $m\simeq 10^{18}$ GeV, and $\phi_i\equiv 2\times 10^{19}$ GeV.}
\label{field}
\end{figure}

The Klein-Gordon equation of the inflaton in the SR approximation is
\begin{equation}
\dot{\phi}\simeq-\frac{V'}{3H}=-\frac{M_4}{\sqrt{24\pi}}\frac{V'}{\sqrt{V}}\simeq-\frac{M_4M^2m^{2n}(n-1)}{\sqrt{6\pi}}\frac{1}{\phi^{2n+1}},\label{17}
\end{equation}
where $M_4$ is the 4D Planck scale. When $\phi>m$ or $\phi/m >1$. The solution of Eq.\ref{17} has the SR form,
\begin{equation}\label{phi(t)4D}
\phi(t)=\left[(2n+2) \left(\frac{2M_4 M^2 m^{2n}(1-n)t}{\sqrt{6 \pi }}+\frac{\phi_i^{2n+2}}{2 (n+1)}\right)\right]^{\frac{1}{2n+2}},
\end{equation}
in which $\phi_i$ is the initial inflaton field. In Figure \ref{field}, it decreases with $t$ but as $n$ increases this decrease becomes less. There is a similarity between Fig. \ref{field} and \ref{potential_figure}, the SR shape is formed. This slow-roll is usually stored in the value of $m$.

The slow-roll parameters are calculated using the derivatives of the potential as follows:
\begin{equation}
\epsilon=\frac{M_4^2}{16\pi}\left(\frac{V'}{V}\right)^2=\frac{M_4^2(n-1)^2 m^{4n}}{4\pi \phi^2\left(m^{2n}+\phi^{2n}\right)^2},
\end{equation}
\begin{align}
\eta=\frac{M_4^2}{8\pi}\left(\frac{V''}{V}\right)&=M_4^2(n-1)m^{2n}\nonumber\\
&\times\frac{\left[(2n-3) m^{2n}-(2n+1) \phi^{2n}\right]}{4\pi\phi^2 \left(m^{2n}+\phi^{2n}\right)^2}.
\end{align}
The condition to end inflation is $|\eta|\simeq 1$, which implies
\begin{equation}\label{phi_f}
\phi_f=\left(\frac{M_4^2}{4\pi}\right)^{1/(2n+2)}\left[(n-1)(2n+1)m^{2n}\right]^{1/(2n+2)}.
\end{equation}
The number of e-folds is
\begin{equation}
N=\frac{8\pi}{M_4^2}\int_{\phi_f}^\phi\frac{V}{V'}d\phi\simeq\frac{2\pi}{M_4^2m^{2n} (n^2-1)}(\phi^{2n+2}-\phi_f^{2n+2}).\label{22N}
\end{equation}
With Eq. \ref{22N} and Eq. \ref{phi_f}, the inflaton field depending on $N$ can be calculated as follows
\begin{equation}\label{phi_i}
\phi(N)=\left[\frac{M_4^2m^{2n}(n-1)[(n+1)N+2n+1]}{2\pi}\right]^{1/(2n+2)}.
\end{equation}
The scalar spectral index is an approximation, namely,
\begin{align}
n_s&=1-6\epsilon+2\eta\\
&=1-\frac{(n-1)M_4^2 m^{2n}\left(n m^{2n}+(2n+1)\phi^{2n}\right)}{2\pi\phi^2\left(m^{2n}+\phi^{2n}\right)^2}\\
&\simeq 1-\frac{(2n+1)}{(n+1)N+2n+1},\label{ns}
\end{align}
where we used Eq. \ref{phi_i}. The tensor-to-scalar ratio is calculated through the following steps:
\begin{align}
r&=16\epsilon\\
&\simeq \frac{4M_4^2(n-1)^2 m^{4n}}{\pi \phi^2\left(m^{2n}+\phi^{2n}\right)^2}\nonumber\\
&=4\left(\frac{\sqrt{\pi}m}{M_4}\right)^{2n/(n+1)}(n-1)^2\nonumber\\
&\times\left[\frac{2}{(n-1)[(n+1)N+2n+1]}\right]^{(2n+1)/(n+1)},\label{r}
\end{align}

The slow-roll parameter depends on the third derivative of $V$,
\begin{equation}
\xi=\frac{M_4^4}{64\pi^2}\frac{V'V'''}{V^2}.\label{epsilon}
\end{equation}
With Eq.\ref{epsilon}, the running spectral index is approximately

\begin{align}
\frac{dn_s}{d\ln k}&=16\epsilon\eta-24\epsilon^2-2\xi\\
&=-M_4^4 (n-1)^2 m^{4 n}\times\nonumber\\ 
&\frac{\left((2 n+1) m^{2 n} \phi ^{2 n}+n m^{4 n}+\left(2 n^2+3 n+1\right) \phi ^{4 n}\right)}{4 \pi ^2 \phi ^4 \left(m^{2 n}+\phi ^{2 n}\right)^4}\nonumber\\
&\simeq -\frac{m^{4n}M_4^4(n-1)^2(1+3n+2n^2)}{4\pi^2}\frac{1}{\phi^{2(2n+2)}}\nonumber\\
&=-\frac{(2n^2+3n+1)}{\left[(n+1)N+2n+1\right]^2}\sim-\frac{2}{N^2}.
\end{align}

With $N\sim 50-60$, the running spectral index is about $\text{a few} \times 10^{-4}$, is very small, thus it is compatible with observations of the Planck Collaboration  \cite{planck}. These results are better than other models. However, $r$ in Eq.\ref{r} is too small compared to observations. This is also the reason why we consider the inflation scenario in RSII model. This will be explained in the section \ref{comparisons}.
	
\section{The inflation in the RSII model}\label{RS2}

The RSII model states that only the (effective) gravitational background is altered and the inflaton field is restricted on the brane. This illustrates how matter fields only exist in the typical 3+1 spacetime (the brane), but gravity may operate in 4+1 spacetime (the bulk). We make a widely held premise that the additional dimension just modifies the effective gravitational background on the brane and neither expands nor contracts. Here, we examine the RSII model's predictions for the inflaton potential (Eq. \ref{potential}). In Ref. \cite{roy} and its references, the overall structure of inflation in the RSII model is covered.

\begin{figure}[h!]
	\centering
	\includegraphics[scale=0.8]{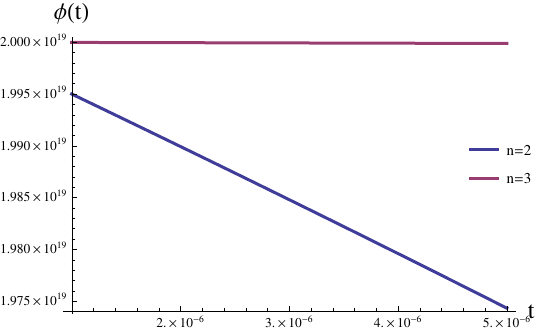}
	\caption{The $\phi(t)$ field in the RSII model with $n=2,3$, $M\simeq 10^{15}$ GeV, $M_4\simeq 10^{19}$ GeV, $m\simeq 10^{18}$ GeV, and $\phi_i\equiv 2\times 10^{19}$ GeV.}
	\label{fieldrs}
\end{figure}

The KG equation in the SR approximation is
\begin{equation}
\dot{\phi}\simeq-\frac{V'}{3H}=-\frac{M_5^3}{4\pi}\frac{V'}{V}\simeq-\frac{M_5^3 (n-1) m^{2n}}{2\pi\phi^{2n+1}},
\end{equation}
where $M_5$ is the five-dimensional Planck scale. The parameter $M_5$ can be thought of as the fundamental energy scale of gravity in the bulk, whereas the usual 4D Planck scale $M_4$ is just an effective energy scale of gravity on the brane. The solution of this equation is
\begin{equation}\label{phi(t)RS2}
\phi(t)=\left[(2n+2) \left(\frac{M_5^3m^{2n}(1-n)t}{2\pi }+\frac{\phi_i^{2n+2}}{2 (n+1)}\right)\right]^{\frac{1}{2n+2}},
\end{equation}
where $\phi_i$ is the initial value of the inflaton field. The behavior of $\phi(t)$ is Fig.\ref{fieldrs}, similar to Fig.\ref{field}, in which the larger $n$ gets, the flatter they become as $t$ increases. 

From the potential Eq. \ref{potential}, the SR parameter is
\begin{align}
\epsilon&=\frac{3M_5^6}{16\pi^2}\frac{V'^2}{V^3}\nonumber\\
&=\frac{3 M_5^6 (n-1)^2 m^{4n} \phi^{-2n} \left(m^{2n}+\phi^{2n}\right)^{-\frac{n+1}{n}}}{4\pi^2 M^4}.
\end{align}
This indicates that it is proportional to $M_5^6/M^4$, this is also similar to $\eta$,
\begin{align}
\eta&=\frac{3M_5^6}{16\pi^2}\frac{V''}{V^2}\nonumber\\
&=\frac{3 M_5^6 (n-1) m^{2 n} \phi ^{-2 n} \left(m^{2 n}+\phi ^{2 n}\right)^{-\frac{n+1}{n}}}{8 \pi ^2 M^4}\nonumber\\
&\times \left((2n-3) m^{2 n}-(2n+1) \phi^{2n}\right).
\end{align}

By using the condition for inflation to end $|\eta|\simeq 1$, the last value of $\phi$ will be, at the end of inflation,
\begin{equation}\label{phi_f_RS2}
\phi_f=\left(\frac{3m^{2n}M_5^6(n-1)(2n+1)}{8M^4\pi^2}\right)^{1/(2n+2)}.
\end{equation}
The number of e-folds ($N$) is
\begin{equation}
N=\frac{-16\pi^2}{3M_5^6}\int_\phi^{\phi_f}\frac{V^2}{V'}d\phi\simeq\frac{4\pi^2M^4(\phi^{2n+2}-\phi_f^{2n+2})}{3M_5^6m^{2n}(n^2-1)}. \label{37}
\end{equation}
From Eqs. \ref{phi_f_RS2} and \ref{37} we will calculate $\phi$ depending on $N$,
\begin{equation}\label{phi_i_RS2}
\phi(N)=\left[\frac{3M_5^6m^{2n}(n-1)[(n+1)N+(2n+1)]}{4\pi^2M^4}\right]^{\frac{1}{2n+2}}.
\end{equation}

With Eq. \ref{phi_i_RS2}, the scalar spectral index is calculated in terms of $\epsilon$ and $\eta$ as follows
\begin{align}
n_s&=1-6\epsilon+2\eta\\
&=1-\frac{3m^{2n}M_5^6(n-1)}{4M^4\pi^2}\left[\frac{m^{2n}(4n-3)+(1+2n)\phi^{2n}}{\phi^{2n}(\phi^{2n}+m^{2n})^{(1+n)/n}}\right]\nonumber\\
&\simeq 1-\frac{2n+1}{(n+1)N+2n+1},\label{ns_RS2}
\end{align}
By using Eq. \ref{phi_i_RS2}, the tensor-to-scalar ratio is proportional to $\epsilon$, we will derive it as a function of $N$,
\begin{align}
r&=24\epsilon\\
&\simeq\frac{18m^{4n}M_5^6(n-1)^2}{M^4\pi^2}\frac{1}{\phi^{2(2n+1)}}\\
&=18\left(\frac{\pi mM^2}{M_5^3}\right)^{2n/(n+1)}(n-1)^2 \nonumber\\
&\times\left[\frac{4}{3(n-1)[(n+1)N+2n+1]}\right]^{(2n+1)/{(n+1)}},\label{r_RS2}
\end{align}
Eq. \ref{ns_RS2} and Eq. \ref{ns} are respectively the scalar spectral indexs $(n_s)$ of the two models; are the same because they simply lost $M_4$ or $M$. 

The tensor-to-scalar ratios $(r)$ in the two models are quite distinctly different as see in Eqs. \ref{r_RS2} and \ref{r}; in RSII model, it depends on $M$ and $M_5$; in the 4D model, it only depends on $M_4$. Therefore, we can surmise that $r$ in the RSII model will be much larger than in the 4D model.

To calculate the running spectral index, we accept another slow-roll parameter as in Eq.\ref{epsilon}. Thus the running spectral index \cite{lidsey} is calculated and its result depends only on $N$, namely,
\begin{align}
\frac{dn_s}{d\ln k}&=16\epsilon\eta-18\epsilon^2-2\xi\\
&= -\frac{9m^{4n}M_5^{12}(n-1)^2}{16M^8\pi^4}\nonumber\\
&\times\bigg(\frac{\left[nm^{4n}(4n-3)+m^{2n}(8n^2-2n-3)\phi^{2n}\right]}{\phi^{4n}(m^{2n}+\phi^{2n})^{(2n+2)/n}}\nonumber\\
&+\frac{\left[(2n^2+3n+1)\phi^{4n}\right]}{\phi^{4n}(m^{2n}+\phi^{2n})^{(2n+2)/n}}\bigg)\\
&\simeq -\frac{9m^{4n}M_5^{12}(n-1)^2(2n^2+3n+1)}{16M^8\pi^4}\frac{1}{\phi^{2(2n+2)}}\\
&=-\frac{(2n^2+3n+1)}{\left[(n+1)N+2n+1\right]^2}\sim-\frac{2}{N^2}.
\end{align}

Indeed the running spectral indices are the same in two models. Another important observable is the amplitude of scalar perturbation (the scalar power spectrum amplitude):
\begin{align}
A_s&\simeq\frac{1024\pi^4}{81M_5^{18}}\frac{V^6}{V'^2}\\
&= \frac{256\pi^4M^{16}}{81M_5^{18}m^{4n}(n-1)^2}\frac{1}{\phi^{6-8n}(m^{2n}+\phi^{2n})^{2-4/n}}\\
&\simeq \frac{256}{81}\left[\frac{\pi^2M^{4(2n+3)}}{(n-1)m^{2n}M_5^{3(2n+4)}}\right]^{1/(n+1)}\nonumber\\
&\times\left[\frac{3((n+1)N+2n+1)}{4}\right]^{(2n+1)/(n+1)}.\label{A_s}
\end{align}

We will discuss the results of the shaft inflation in the RSII model in the next section.

\section{The fitting with experiments}\label{comparisons}

The evolution of the inflaton field in 4D spacetime (Eq. \ref{phi(t)4D}) and in the RSII model (Eq. \ref{phi(t)RS2}) is shown in Figure \ref{field} and \ref{fieldrs}. They both exhibit similar shapes with respect to $t$.

\begin{figure}[h!]
	\centering
	\includegraphics[scale=0.9]{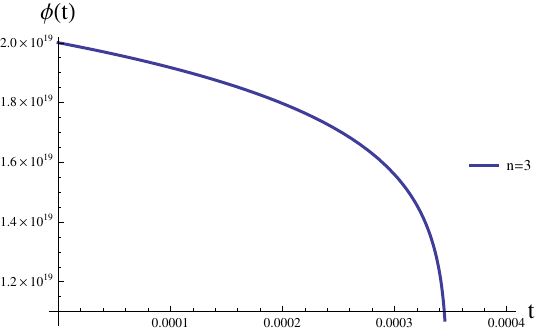}
	\caption{The $\phi(t)$ field in the RSII model with $n=3$, $M\simeq 10^{15}$ GeV, $M_4\simeq 10^{19}$ GeV, $m\simeq 10^{18}$ GeV, and $\phi_i\equiv 2\times 10^{19}$ GeV. The slope increases very rapidly with $t$.}
	\label{fieldrs2}
\end{figure}

\begin{figure}[hbtp]
\centering
\includegraphics[scale=0.7]{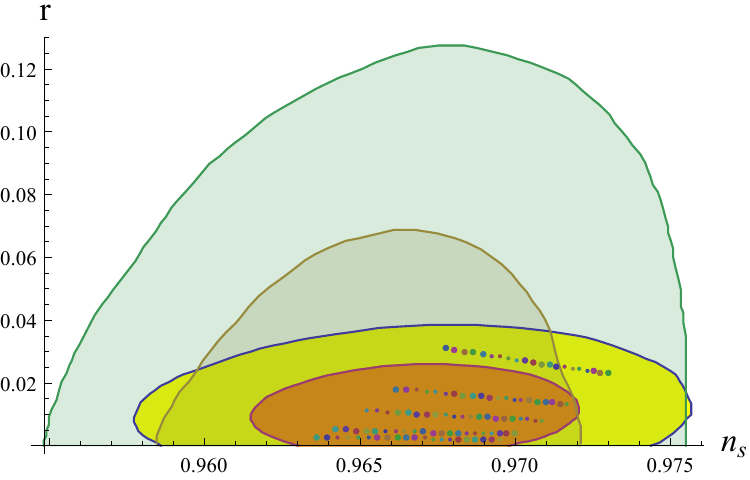}
\caption{Compatibility between experimental and calculated values in the RSII model. $N$ from $50$ to $60$, on each dot line, from the first dot on the left to the last dot. From top to bottom, the dot lines correspond to $n=2,3,4,6,8$, respectively. $M \simeq 10^{15}$ GeV and $M_5\sim 1.4\times 10^{16}$ GeV. The yellow and orange areas are the BICEP2 results \cite{bicep}. The light blue areas are the Planck results \cite{bicep,planck1,planck}.}
\label{planck_figure}
\end{figure}

There is a similarity between Fig. \ref{field} and \ref{fieldrs}, but $\phi(t)$ in Fig. \ref{fieldrs} shows a significant difference between $n=2$ and $n=3$ compared to Figure \ref{field}. This is also clearly shown in Fig. \ref{planck_figure} (when examining the values of ($n_s$,$r$) in the RSII model with values of $n$).

Furthermore, from Fig. \ref{fieldrs2}, the slope of the field is very large. This is also consistent in the slow-roll scenario in the shaft inflation.

\begin{figure}[h!]
	\centering
	\includegraphics[scale=0.9]{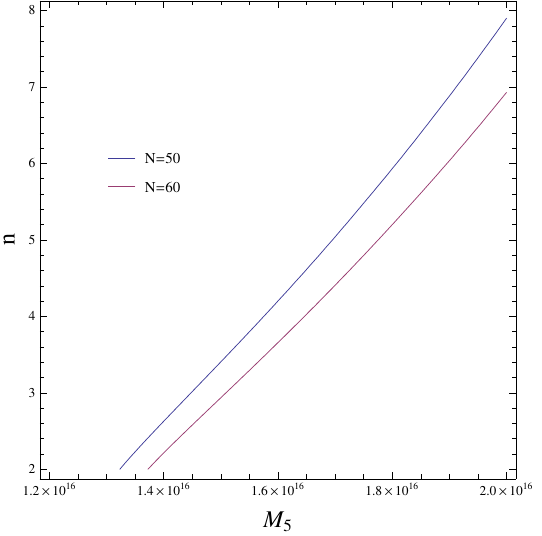}
	\caption{$M_5$ in the RSII model depends on $n$ with $A_s=2\times 10^{-9}$.}
	\label{field6a}
\end{figure}

In Figure \ref{field6a}, the lines with different $N$ values are quite similar, so $N$ only affects the magnitude of $n_s$ and $r$ but does not affect the variation of these quantities. Furthermore, as $N$ increases, $M_5$ tends to increase, and $n$ decreases. This suggests that if the greater the expansion of the universe is during the inflationary phase, the larger $M_5$ must also be.

\begin{figure}[h!]
	\centering
	\includegraphics[scale=0.9]{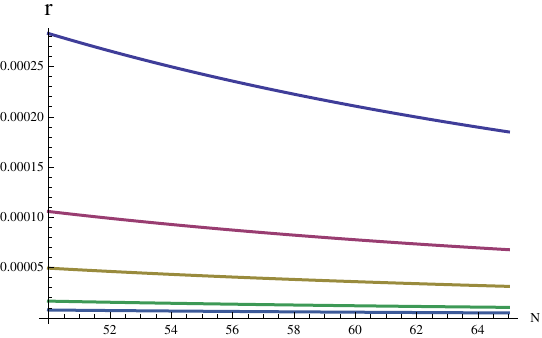}
	\caption{The $r$ function depends on $N$ in the 4D model. From top to bottom, the lines correspond to $n=2,3,4,6,8$, respectively.}
	\label{field6}
\end{figure}

\begin{figure}[h!]
	\centering
	\includegraphics[scale=0.9]{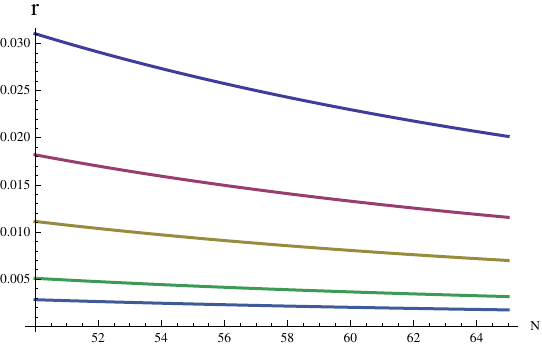}
	\caption{The $r$ function depends on $N$ in the RSII model. From top to bottom, the lines correspond to $n=2,3,4,6,8$, respectively.}
	\label{field5}
\end{figure}

Now that we compare the observables of two models with each other and with observations. The typical energy scale of inflation is $M\simeq 10^{15}$ GeV and assuming that $m\simeq 10^{18}$ GeV, we can find the value of $M_5$ by demanding that the results in Eq. \ref{A_s} match the observed value $A_s^{Planck}\sim 2\times 10^{-9}$ \cite{planck}. In Fig.\ref{field6a}, it is found that $M_5$ is typically of order $M_5\sim [1.4-2]\times 10^{16}$ GeV. Based on the results in Fig. \ref{field6} and Fig. \ref{field5}, it can be seen that the RSII model predicts larger tensor-to-scalar ratio than the 4D case.

The average value of $r$ in the standard 4D model is approximately $10^{-4}$. Nevertheless, in the RSII model, this value is $10^{-2}$.

\begin{figure}[h!]
	\centering
	\includegraphics[scale=0.95]{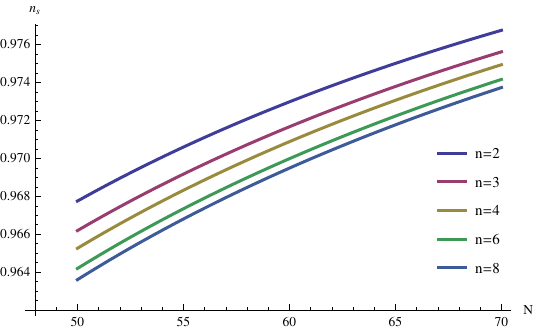}
	\caption{The $n_s$ function depends on $N$.}
	\label{field4}
\end{figure}

$n_s$ increases almost linearly with $N$ as shown in Fig. \ref{field4}. We can approximate as follows: $n_s=0.006N$. The average value of $n_s$ is approximately $0.97$.

In Figs. \ref{field5} and \ref{field6}, $r$ decreases slightly as $n_s$ increases. When $n>4$, the values of $r$ are quite close to each other, whereas for $n<4$, $r$ changes significantly as $n$ increases. In particular, when $n$ increases from 2 to 3, $r$ changes very strongly. This is different from the shaft inflation.

$r$ in the RSII model is about 100 times larger than in the 4D model as seen in Figures \ref{field6} and \ref{field5}. Therefore, $r$ in the RSII model fits experiment better. This is also the first basis for asserting that the RSII model is better than the standard 4D model.

However, their behavior with respect to $N$ is the same, this is similar to that in Ref. \cite{EPJC}. $n_s$ in both models lies in the central range of the experiment, that is, from $0.964$ to $0.972$.

As $n$ increases from $2$ to $8$, $r$ decreases by a factor of 10. From Fig. \ref{field5}, with just $n=2,3$ the result $r$ matches the experimental data very well. The value of $r$ is always less than $0.03$.

In Figures \ref{field6}, \ref{field5}, and \ref{field4}, the larger $n$ is, the closer the lines are to each other. In other words, a clear hierarchy of values for the quantities ($r, n_s$) only occurs with small $n$. We need to combine the data from the pair ($r, n_s$) to compare with experimental results. However, since the $r$ in the RSII model is better than the standard 4D model, we only need to use the pair ($r, n_s$) in the RSII model to compare with experimental results, as shown in Fig. \ref{planck_figure}. 

In most models, the larger $n$ is, the smaller $r$ becomes, although $n$ cannot be much larger than 2. In our model, $n$ only needs to be $2$ or $3$, yielding results consistent with experimental data. This is an improvement in scalar potentials aimed at reducing the exponent.

Fig. \ref{planck_figure} compares the observations in Ref. \cite{bicep} with the predictions of the RSII model. For a range of $n$ values, we see that the findings are in great agreement with observation. The RSII model forecasts higher tensor-to-scalar ratios, which can be seen in upcoming accurate measurements, in contrast to the 4D scenario.

In Fig. \ref{planck_figure}, the lines $r$ corresponding to $n>2$ are fairly uniform and quite far from the line with $n=2$. This is analogous to the lines of $\phi(t)$ according to $n$.

\section{Conclusions and discussions}\label{conclu}

In this article, the inflation has been calculated in the RSII and the typical 4D model. The outcomes were then contrasted with observations. We have attempted to provide a context to explain the inflationary potential from the dilaton potential in the 2T Physics. This is clearly demonstrated in Sec.\ref{2T} and Sec.\ref{4D}. It is similar to explaining other inflationary potentials from quantum superstring theory or supersymmetry.

Particularly in the case where $n=3$, $r<0.03$ is at a $95\%$ confidence level, yet it is still quite large for experimental confirmation (as shown in Fig. \ref{planck_figure}). Furthermore, as assessed in the previous section, the scalar potential in this case is consistent with the dilaton potential in the 2T physics. 

The results in this paper have two implications: firstly, they suggest an explanation of the inflaton field, thereby demonstrating consistency in explaining many current problems using the dilaton field, such as the matter-antimatter asymmetry problem combined with the inflation problem; secondly, they present the inflation scenario that is modified from the shaft inflation and it yields better results compared to experiments when the order $n$ is small.

We got an intuitively predicted feature from the dynamical perspective: the RSII model's inflaton field rolls more slowly because of the Friedman equation's adjustment during inflation. In particular, the Hubble rate squared increases friction in the Klein-Gordon equation. The RSII model's scale factor grows quickly, making it easy to produce enough e-folds to account for the flatness and horizon issues. The predictions of the inflation in the RSII model are in great accord with observation, especially for small $n$ (because $n$ is much larger than $3$, $r$ will be very small; experimentally, this is very difficult to determine). $M_5\simeq 10^{16}$ GeV is larger than $10^9$ GeV \cite{EPJC}, this is consistent with the predictions. It is a significant prediction that may be used to investigate the implications of the additional dimension in different situations. Furthermore we find that this scenario can be extended to investigate in other models such as the DGP model \cite{dgp}.

The results of this article will be a suggestion for investigating other problems such as describing the accelerating expansion of the universe with the gravitational actions in 2T as in Refs.\cite{g2t,g2t2}. From there, understand more about the dilaton (as a candidate for dark matter or dark energy) and the time dimension. 

One of the effects that we can see right away is the direct interaction between the dilaton and the gravitational field, as seen from the actions in Refs.\cite{g2t,g2t2}. These interactions involve energy scales in dimensions higher than $4$, and also cause the Friedmann equations to be slightly modified from their usual form. The modifications are all stored in the 5-dimensional action of the RSII model. In other words, the RSII model can be considered as a shadow of the 2T model. 

The 2T model contains the dilaton field, which gives us a natural explanation of the dilaton field in the context of gravity, and the dilaton can be a candidate for the inflaton.

\begin{figure}[h!]
	\centering
	\includegraphics[scale=0.5]{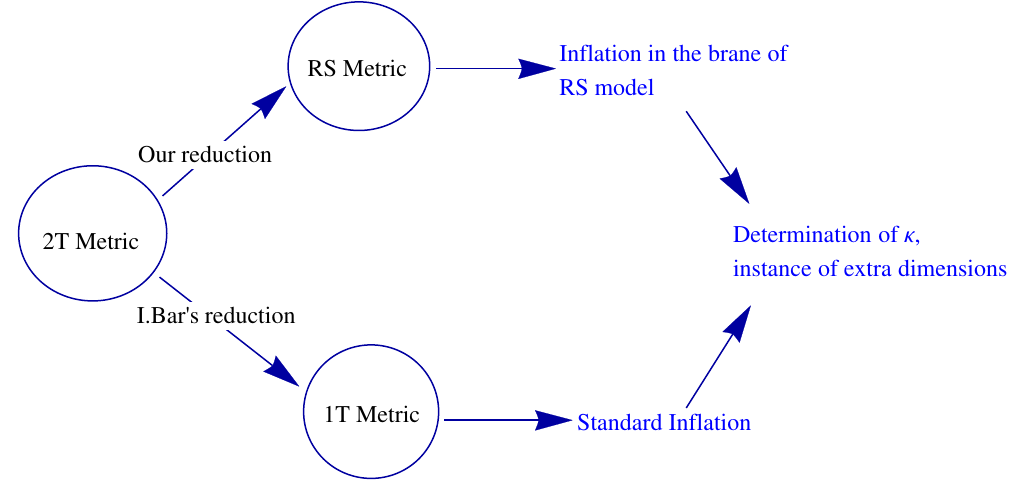}
	\caption{The inflation survey chart in the article. The final step in the chart (finding the value of $\kappa$) is one of the things we're aiming for in the future.}	\label{findkappa}
\end{figure}

We also summarize the inflation survey process as shown in Fig. \ref{findkappa}. The survey results show that the RSII model with this new potential fits empirical data better than the standard 4D model. Furthermore, according to Fig. \ref{findkappa}, and the results in this article, a future task is to search for conditions or values for the parameter $\kappa$. This will lead to an explanation of the direct or indirect existence of the extra dimension.

Note that Fig. \ref{findkappa} only shows the direction in using inflation data to determine $\kappa$. This can be generalized as follows: once we determine the scalar field functions during the inflation period, from Eq. \ref{11}, we can estimate $\kappa$. However, we still do not have enough additional basis to calculate $\kappa$; this is just a suggestion worth considering.

We summarize the entire paper by stating that, with a modified form of the shaft inflation potential, compatible with all exponents $n$, the RSII model is better at examining the inflation than the standard 4D model and originates from the dilaton potential in the 2T Physics. This also shows a way to determine the extra dimensions through inflation data.

\section*{Acknowledgment}

This research is funded by University of Science, VNU-HCM under grant number T2025-51.\\

This article celebrates the 15th anniversary of the Cosmology direction established at the Department of Theoretical Physics, University of Science, Vietnam National University, Ho Chi Minh City. It is also in memory of my teacher, Dr. Vo Thanh Van.

\end{document}